
\input harvmac
\input epsf
\input tables
\def\NP#1{Nucl.~Phys. {\bf B{#1}}}
\def\PL#1{Phys.~Lett. {\bf B{#1}}}
\def\PREP#1{Phys.~Rep. {\bf {#1}}}

\def\PRD#1{Phys.~Rev. {\bf D{#1}}}
\def\PTP#1{Prog.~Theor.~Phys. {\bf {#1}}}
\def\PRL#1{Phys.~Rev.~Lett. {\bf {#1}}}
\catcode `\@=11
\newfam\mibfam

\font\tenmib=cmmib10
\font\sevenmib=cmmib7
\font\fivemib=cmmib5
\skewchar\tenmib='177
\skewchar\sevenmib='177
\skewchar\fivemib='177
\textfont\mibfam=\tenmib
\scriptfont\mibfam=\sevenmib
\scriptscriptfont\mibfam=\fivemib
\let\rel@x=\relax
\let\n@expand=\relax
\newcount\f@ntkey  \f@ntkey=0
\def\f@m{\afterassignment\samef@nt\f@ntkey=}
\def\samef@nt{\fam=\f@ntkey \the\textfont\f@ntkey\rel@x}
\def\rm{\n@expand\f@m0 }
\def\mit{\n@expand\f@m1 }
\def\cal{\n@expand\f@m2 }
\def\it{\n@expand\f@m\itfam}
\def\sl{\n@expand\f@m\slfam}
\def\bf{\n@expand\f@m\bffam}
\def\tt{\n@expand\f@m\ttfam}
\def\mib{\n@expand\f@m\mibfam}
\def\lsim{\ \raise.3ex\hbox{$<$\kern-.75em\lower1ex\hbox{$\sim$}}\ }
\def\gsim{\ \raise.3ex\hbox{$>$\kern-.75em\lower1ex\hbox{$\sim$}}\ }
\def\gl{\ \raise.5ex\hbox{$>$}\kern-.8em\lower.5ex\hbox{$<$}\ }

\catcode `\@=12
%
\Title{\vbox{\baselineskip12pt\hbox{KYUSHU-HET-10}\hbox{SAGA-HE-50}}}
{\vbox{\centerline{The Effective Potential of Electroweak Theory}
\vskip2pt
       \centerline{with Two Massless Higgs Doublets}
\vskip2pt
       \centerline{at Finite Temperature}}}
\centerline{Koichi
Funakubo\footnote{$^\star$}{funakubo@sagagw.cc.saga-u.ac.jp},
Akira Kakuto$^1$\footnote{$^\dagger$}{kakuto@fuk.kindai.ac.jp}
and Kazunori
Takenaga$^2$\footnote{$^\ddagger$}{f77498a@kyu-cc.cc.kyushu-u.ac.jp}}
\bigskip
\centerline{\it Department of Physics, Saga University, Saga, 840 JAPAN}
\smallskip
\centerline{$^1${\it Department of Liberal Arts, Kinki
University in Kyushu, Iizuka, 820 JAPAN}}
\smallskip
\centerline{$^2${\it Department of Physics, Kyushu
University, Fukuoka, 812 JAPAN}}
%
%
\vskip 1cm
The effective potential of electroweak theory with two
massless Higgs doublets at finite temperature is studied.
We investigate phase structure and critical temperature
in this model by numerical analysis without high-temperature
expansion. The phase transition is found to be of first order.
The critical temperature is shown to be relatively low for
typical scalar masses. The free energy of the critical bubble
is calculated with some approximations and we find that 
the bubble nucleation can occur at a temperature a little
below the critical temperature. We also discuss the
possibility of the electroweak baryogenesis.
\Date{10/93}
%
\newsec{Introduction}
Recently, much attention has been paid to baryogenesis at
the electroweak phase transition (EWPT).
In addition to the great success of low-energy phenomenology
in the standard model, it is well known that the standard
model may satisfy the famous \hbox{Zakharov's} three
conditions\ref\za{A.~D.~Zakharov, JETP Lett. ${\bf 5}$ (1967) 24.}
to generate baryon number : (i) There are
processes which violate baryon number. (ii) Underlying
dynamics involves CP violating processes. (iii) There exist
non-equilibrium processes which are caused by first-order
phase transition. There has been a lot of studies to try to
generate baryon number at the EWPT to explain the baryon asymmetry
of the Universe observed today. Whether the condition (iii) is
satisfied by underlying dynamics or not is usually determined by
analyzing the effective potential at finite temperature. In the
standard model, the negative tree-level mass term of the Higgs field
causes a serious infrared problem that the effective potential
becomes complex at one-loop level. Thus it is very difficult
to obtain a reliable result by perturbation theory.
One usually resorts to high-temperature expansion and takes
terms including $O(M^3T)$ term in bosonic contributions to the
effective potential\foot%
{For fermions there is no $O(M^3T)$ term.}.
The $O(M^3T)$ term is very crucial for the first-order
phase transition. However this term is complex when the order
parameter takes small values.
Many works have been performed to improve the potential at finite
temperature\ref\tz{N.~Turok and J.~Zadrozny, \NP{369} (1992) 729.}%
\ref\ca{M.~Carrington, \PRD{45} (1992) 2933.}.
It is suggested that the phase transition is weakly first order
for relatively light Higgs boson in some recent works\ref\ba
{W.~Buchm\"uller, Z.~Forder,T.~Helbag and D.~Walliser,
DESY 93-021(1993).}. In order to preserve baryon number
after finishing EWPT, the rate of anomalous process induced
by the sphaleron must be smaller than that of the expansion of
the Universe. This, in turn, is converted into a strong constraint
on Higgs boson mass.  It leads us to an upper bound of the Higgs
boson mass, $M_H=35 \sim 50$ GeV. On the other hand, the lower
bound of the Higgs boson mass found by experiments at LEP is about
60 GeV. Moreover, it is believed that the CP violation in
Kobayashi-Maskawa scheme of the standard model is too feeble
to generate baryon asymmetry of the Universe observed today.\par
Two-Higgs-doublet model has recently been investigated because of
possibilities to avoid these difficulties\tz. In
two-Higgs-doublet model, there could appear a new CP violating phase
in the Higgs sector, and it is expected that this new CP violation
could make baryon density large enough to be comparable with the
observed value. Moreover it is argued that the constraint on the
lightest Higgs boson mass in order not to wash out baryon density
after EWPT is not so stringent and the Higgs boson could be as
heavy as about $100$ GeV \tz, which is consistent with
experiments. In general there are three order
parameters\foot{Assuming the $U(1)_{em}$ invariance.}
in this model, and these parameters are relevant for both
baryogenesis and new CP violation in weak interactions. It is
complicated to analyze the effective potential including one-loop
effects at finite temperature, and to find the minimum of the
potential to study whether the new CP violation is realized or not.
Difficulties are mainly due to mixing mass terms of the two Higgs
doublets if one needs to include a possibility of the new CP
violation.  Furthermore, as in the usual standard model, there
appears a serious infrared problem associated with negative
tree-level mass terms of Higgs fields. We believe that many
analyses should be worked out to understand the two-Higgs-doublet
model with negative mass terms.\par
In this paper we shall investigate phase structure of the electroweak
theory with two massless Higgs doublets. This model is much simpler
than the two-doublet model with tree-level mass terms. There is no
infrared difficulty and it is easy to analyze the effective potential
at finite temperature. We shall study critical temperature and the
order of EWPT by analyzing the one-loop effective potential at finite
temperature without high-temperature expansion. In fact it turns out
that high-temperature expansion cannot be applied at EWPT period in
this model. The order of phase transition is found to be first order
for wide ranges of scalar masses. At very high temperatures, loop
expansion of the effective potential might be problematic
\ref\dj{L.~Dolan and R.~Jakiw, \PRD{9} (1974) 3320.}.
In our model, the critical temperatures are found to be significantly
low compared with those in the standard model for wide ranges of scalar
masses. This fact suggets that our calculations may be reliable at this
range of temperature. We will find that the rate of the anomalous
process after EWPT is sufficiently slow in order not to wash out
baryon density. We obtain a critical temperature $T_C \simeq 100$
GeV for relatively heavy scalar masses\foot%
{ This $T_C$ corresponds to the scalar mass set II defined in
section 2.}. It is expected these features of the phase transition
in this model are also shared in the model with small tree-level
masses of Higgs fields. The phase transition in the supersymmetric
standard model may also have the same features if the tree-level
Higgs masses are small. We believe that our model can provide many
attractive features for generation of baryon number at the EWPT,
though new CP violating phase in the Higgs sector cannot be introduced.
\par
In the next section we give the one-loop effective potential at
finite temperature in the present model. In section 3 we study the
effective potential for various sets of scalar mass parameters and
find the critical temperature for each case by numerical analysis.
The EWPT is found to be of first order. We also study briefly the
bubble nucleation below the critical temperature.
Section 4 is devoted to conclusions and discussions.
\newsec{The Effective Potential at Finite Temperature }
In imaginary time formalism, the effective potential is automatically
separated into zero-temperature and finite-temperature
parts\dj.
In the case of one-Higgs-doublet model, we are afflicted with a serious
infrared problem in temperature-dependent part, which is
due to negative tree-level mass term of the Higgs field as mentioned
before. In the present model, however, there is no such
kind of problem, because tree-level Higgs mass terms are absent
by assumption. It is not necessary to use high-temperature
expansion in order to study the structure of phase
transition. As we will see later, high-temperature expansion
should not be adopted in this model, because it turns out
that the critical temperature is low compared with the mass of the
heaviest scalar. The zero-temperature part of the effective
potential was analyzed in detail by Inoue, Nakano and one of the
authors \hbox{(A.K.)\ref\ikn{K.~Inoue, A.~Kakuto and Y.~Nakano,
\PTP{63} (1980) 234.}.} We will start with a review on the
zero-temperature effective potential in the massless
two-Higgs-doublet model in the next subsection.
\subsec{Model and one-loop potential at zero temperature}
We work in the usual $SU(2)\times U(1)$ gauge theory of the electroweak
model with two massless Higgs doublets and with $N_g$ generations
of quarks and leptons. The effect of $SU(3)_c$ color strong
interactions appears only through quark loops in the effective
potential. The number of colors is denoted by $N_c(=3$ for $SU(3)_c$).
Quantum number assignment of $SU(2) \times U(1)$ for quarks and
leptons are as follows:
$$
Q_{LA} \equiv {u_A \choose d_A}_L;  \quad Y(Q_{LA})={1\over 6},\quad
l_{LA} \equiv {\nu_A \choose e_A}_L; \quad Y(l_{LA})=-{1\over 2},
$$
$$
u_{RA};\quad Y(u_{RA})={2\over 3},\quad d_{RA};\quad Y(d_{RA})=-{1\over 3},
\quad e_{RA};\quad Y(e_{RA})=-1,
$$
where $Y$ stands for the $U(1)$ hypercharge and $A$ distinguishes
generations. We introduce two massless Higgs doublets
$\Phi_1$ and $\Phi_2$ with $Y(\Phi_1)=Y(\Phi_2)=1/2$. In order to
avoid dangerous tree-level flavor-changing neutral interactions
mediated by scalar exchange, we restrict allowed Yukawa
couplings by imposing the following discrete
symmetry\ref\wbrg{S.~Weinberg, \PRL{37} (1976) 657.}, by which only
one kind of Higgs fields interacts with $u_R$ or $d_R$:
\eqn\aa{
\Phi_2 \rightarrow -\Phi_2, \qquad u_{RA} \rightarrow -u_{RA}.
}
Then we have the following Yukawa couplings:
\eqn\ab{
\CL_Y = - \sum_{A,B}^{N_g}\big[{\bar Q_{LA}} {f_{AB}^{(u)}}{\tilde \Phi}_2
{u_{RB}}+ {\bar Q_{LA}} {f_{AB}^{(d)}} {\Phi_1 }{d_{RB}}+
{\bar l}_{LA} f_{AB}^{(l)}\Phi_1 e_{RB} +{\rm h.c.} \big]
,}
where ${\tilde\Phi}_2=i{\tau}^2 \Phi^*_2$ , with $ \tau^a (a=1,2,3)$
being Pauli matrices. The most general renormalizable and
$SU(2)\times U(1)$ invariant Higgs quartic interactions
contain seven types of independent couplings. The discrete symmetry
\aa\  must also be imposed on the Higgs self-interactions,
otherwise the Yukawa couplings \ab\  lose their
meaning\foot{Discussion is given in ref. \ikn.}.
Then the Higgs potential is written as
\eqn\ac{\eqalign{
V_{H}&={\lambda_1 \over 2}({\Phi_1}^\dagger \Phi_1)^2+
{\lambda_2 \over 2}({\Phi_2}^\dagger \Phi_2)^2+
\lambda_3 ({\Phi_1}^\dagger \Phi_1)({\Phi_2}^\dagger \Phi_2)\cr
&+\lambda_4 ({\Phi_1}^\dagger \Phi_2)({\Phi_2}^\dagger \Phi_1)+
{\lambda_5 \over 2}[({\Phi_1}^\dagger \Phi_2)^2+{\rm h.c.}] \quad
(\lambda_{1,2}
>0).\cr
}
}
The vacuum expectation values (VEV) of the Higgs fields are determined
by minimizing the effective potential and their norm is fixed by
$v^2=1/{{\sqrt 2}G_F}$ to describe low-energy phenomena. Here
$G_F=1.16639 \times {10^{-5}}({\rm GeV})^{-2}$
is the Fermi coupling constant. In our model the Higgs potential is a
homogeneous polynomial of Higgs fields. Thus spontaneous symmetry
breaking does not occur at the tree level. We must consider
at least the one-loop potential.\par
Calculating the contribution from one-loop order to a potential 
in a model with more than one Higgs doublets is not so
straightforward as it appears. This is because in multi-doublet
case one cannot choose a renormalization scale such that all the
quartic couplings are small%
\ref\sher{M.~Sher, \PREP{179} (1989) 273.}.
A comprehensive study of the effective potential in multi-doublet
model was carried out by Gildner and Weinberg\foot%
{Simple discussion is also given in ref. \sher.}
\ref\gw{E.~Gildner and S.~Weinberg, \PRD{15} (1976) 3333.}.
We follow their arguments here. First, by use of $SU(2)\times U(1)$
gauge degrees of freedom, we write VEV of the Higgs fields as
\eqn\ad{
\vev{\Phi_1}={\rho \over {\sqrt 2}}{0 \choose n_1}, \quad
\vev{\Phi_2}={\rho \over {\sqrt 2}}{n_4 \choose {n_2+in_3}},
}
where $\rho >0$, $n_i\,(i=1\sim 4)$ are real and $\sum_{i=1}^4 n_i^2 =1$.
\hfill\break
Then VEV of the tree-level potential becomes
\eqn\ae{\eqalign{
V_H^{Tree}={\rho^4 \over 8}\big[\lambda_1 n_1^4+ \lambda_2
(n_2^2+n_3^2+n_4^2)^2 &+
2\lambda_3 n_1^2 (n_2^2+n_3^2+n_4^2) \cr
+ 2\lambda_4 n_1^2 (n_2^2+n_3^2) &+
2\lambda_5 n_1^2 (n_2^2-n_3^2) \big].\cr
}
}
We set a condition that the tree-level potential has minimum value
of zero on some ray $\rho{\mib n} = \rho{\mib n}_0$ for arbitrary
positive $\rho$. The condition implies that
\eqn\af{
{\del \over {\del n_i}}\Bigl({V_H^{Tree} \over \rho^4}\Bigr)=0 \quad
{\rm for}\quad  i=1\sim 4.
}
A solution of \af\ specifies a direction ${\mib n}\equiv
(n_1,n_2,n_3,n_4)$ of VEV, where the tree-level potential takes
minimum value of zero. The solution that breaks the $SU(2)\times U(1)$
symmetry and keeps the $U(1)_{em}$ invariance is uniquely determined as
\foot{We choose the sign of $\lambda_5$ to be negative.
It can be arbitrary chosen by a phase convention of the Higgs field.}
\eqn\ag{
\vev{\Phi_1}={\rho \over {\sqrt 2}}{0 \choose n_{01}}, \quad
\vev{\Phi_2}={\rho \over {\sqrt 2}}{0 \choose n_{02}},
}
\eqn\ahh{
{\rm with}\quad  \lambda \equiv {\sqrt{\lambda_1
 \lambda_2}}+\lambda_3+\lambda_4+\lambda_5=0,
}
where $n_{01}$ and $n_{02}$ are defined by
$$
n_{01}^2\equiv {\sqrt \lambda_2 \over{\sqrt \lambda_1 + \sqrt
\lambda_2}},\quad
n_{02}^2\equiv {\sqrt \lambda_1 \over{\sqrt \lambda_1 + \sqrt \lambda_2}}.
$$
The condition \ahh\ guarantees the minimum value of the tree-level
potential to be zero. In this model there is no new CP violating phase in the
Higgs sector.  As Gildner and Weinberg pointed out,
perturbative calculations are reliable along the ray
$\rho {\mib n}_0$ and the minimum of the one-loop potential
is the deepest one compared with those along other directions.
We assume that the tree-level relation, $\lambda=0$ is
still valid at one-loop order. This means that we choose a
renormalization point $M_R$ at which $\lambda(M_R)=0$.\par
Following the usual procedure, the one-loop potential,
$V_1$ at zero
temperature in
$R_{\xi}$ gauge is given as
\eqn\ah{\eqalign{
V_1={1\over {64 \pi^2}}&\biggl[{3 \over  8}g^4+{3 \over 16}(g^2+g'^2)^2+
{1\over 2}(\sqrt{\lambda_1 \lambda_2}+\lambda_3)^2 +\lambda_1 \lambda_2+
\lambda_5^2 \cr
 &-{\lambda_2 \over{(\sqrt \lambda_1+\sqrt \lambda_2)^2}}\sum_A(N_c
         f_A^{(d)4}+f_A^{(l)4})\cr
 &-{\lambda_1 \over{(\sqrt \lambda_1+\sqrt \lambda_2)^2}}\sum_A N_c
f_A^{(u)4}\biggr]\cr
 &\times
 \rho^4 \biggl({\rm ln}{\rho^2 \over M_R^2}-{25\over 6}\biggr), \cr
}
}
where the renormalization point is defined by
$$
\biggl[{{d^4 V_1}\over{d \rho^4}} \biggr]_{\rho=M_R}=0.
$$
We denote $SU(2)$, $U(1)$ and diagonalized Yukawa
couplings by $g$, $g'$ and $f_A^{(u,d,l)}$  respectively.
Note that the gauge-dependent terms vanish in the
Gilder-Weinberg method. As mentioned before, VEV is determined by
the minimum\foot{The coefficient of $\rho^4
({\rm ln}{\rho^2 \over M_R^2}-{25\over 6})$ must be positive for
$V_1$ to be lower bounded.} of $V_1$, at which it is given by
$$
\rho^2 = v^2=M_R^2{\rm e}^{11/3}.
$$
Eliminating $M_R$ in \ah\  and rescaling $\rho$ as $\rho /v \equiv
\ph$, we have
\eqn\ai{
V_1={1\over {64 \pi^2}}B \ph^4
({\rm ln}\ph^2 -{1 \over 2}),
}
where
\eqn\aj{\eqalign{
B&\equiv 6M_W^4(v)+3M_Z^4(v)+2M_{H^{\pm}}^4(v)+M_h^4(v)+M_A^4(v) \cr
&-4\sum_A\big(N_c M_A^{(u)4}(v)+N_c M_A^{(d)4}(v)+M_A^{(l)4}(v)\big).\cr
}
}
Here various kinds of particle masses are defined as
\eqn\ak{\eqalign{
M_W^2(v)&\equiv {g^2 \over4}v^2, \cr
M_{H^{\pm}}^2(v) &\equiv {1\over 2}({\sqrt{\lambda_1
\lambda_2}}+\lambda_3)v^2, \cr
M_A^2(v) &\equiv -\lambda_5 v^2, \cr
M_A^{(d)2}(v) &\equiv {\sqrt \lambda_2 \over
{\sqrt \lambda_1+\sqrt \lambda_2}}f_A^{(d)2}v^2, \cr}
\quad\eqalign{
M_Z^2(v) &\equiv {{g^2 +g'^2}\over4}v^2, \cr
M_h^2(v)  &\equiv {\sqrt{\lambda_1 \lambda_2}}v^2, \cr
M_A^{(u)2}(v)  &\equiv {\sqrt \lambda_1 \over
{\sqrt \lambda_1+\sqrt \lambda_2}}f_A^{(u)2}v^2, \cr
M_A^{(l)2}(v) &\equiv {\sqrt \lambda_2 \over
{\sqrt \lambda_1+\sqrt \lambda_2}}f_A^{(l)2}v^2. \cr}
}
The scalon, which is a particle associated with scale invariance
in this model, becomes massive after spontaneous symmetry breaking.
The scalon mass, $M_S$ is defined by the inverse propagator evaluated
at zero momentum and is given by \gw.
\eqn\al{\eqalign{
M_S^2&=\biggl[{d^2 V_1 \over d\rho^2}\biggr]_{\rho=v} \cr
     &={G_F \over {4 {\sqrt
2}\pi^2}}\biggl[6M_W^4(v)+3M_Z^4(v)+2M_{H^{\pm}}^4(v)+
                      M_h^4(v)+M_A^4(v)       \cr
    &-4\sum_A M_A^{(l)4}(v)-4N_c \sum_A M_A^{(u)4}(v)
     -4N_c \sum_A M_A^{(d)4}(v) \biggr].\cr
}
}
\par
Here we give sample sets of particle masses, which are used
in later numerical analysis. Recent precision electroweak measurements
of various observables give us important information on the top
quark mass $M_t$.  It is known, by comparing precision measurements
with higher-order corrections in the standard model,
that $M_t \simeq 131^{+47}_{-28}$ GeV \ref\paul{
P.~Langacker, Lectures presented at TASI-92, Boulder, June 1992.}
for the Higgs mass range \hbox{60GeV--1000GeV.}
We assume $M_t = 140$ GeV throughout our analysis.
The other fermion masses are neglected in numerical
calculations. On the other hand, masses of Higgs bosons are not so
strictly constrained by experiments due to their feeble
contributions to electroweak observables. In our model there are four
kinds of scalar particles, charged Higgs $M_{H^{\pm}}$, CP odd
Higgs $M_A$, neutral Higgs (CP even) $M_h$ ,  and scalon $M_S$. 
All these masses are free parameters\foot%
{$M_S$ is determined from \al$\,$ once we fix $M_{H^{\pm}}, M_A,
M_h$ and $M_t$.}. These scalars have not been detected at
laboratories but effects of these particles can enter into the
electroweak observables, which are well measured now, through loop
diagrams \foot{We are going to analyze effects of
these new particles on electroweak observables in forthcoming paper.}.
The scalon can be regarded as the usual physical Higgs boson in the
standard model. Thus it must be heavier than about 60 GeV.  In
this paper we take the following two sets of scalar masses for
illustrative calculations:\hfill\break
$$
{\bf {\rm set\quad I}} \quad  (M_{H^{\pm}},M_A,M_h:M_S)=(350,250,200:80.95)
\,{\rm
GeV},
$$
$$
{\bf {\rm set\quad II}} \quad  (M_{H^{\pm}},M_A,M_h:M_S)=(450,350,250:142.19)
\,{\rm
GeV}.
$$
For the above scalar masses all the quartic couplings are within
perturbative ranges\foot{We assume that $\lambda_1$
and $\lambda_2$ are the same order.}:
$$
0.04<{|\lambda_i|\over {4\pi}}<0.23\,\,\,(i=1\sim 5)\quad
{\rm for\,\,\, set\,\,\, I},
$$
$$
0.08<{|\lambda_i|\over {4\pi}}<0.45\,\,\,(i=1\sim 5)\quad
{\rm for\,\,\, set\,\,\, II}.
$$
\subsec{One-loop effective potential at finite temperature}
The finite-temperature part of the effective potential is given by the
following prescription\dj ,
$$
\eqalign{
\int {{d^4k}\over{(2\pi)^4}}  &\longrightarrow
\int_k \equiv {i T} \sum_n \int {{d^3
{\mib k}} \over
{ (2 \pi)^3}}\quad , \cr
       k_0 &\longrightarrow
\omega_n \equiv \cases{
    {2 \pi n }{i T}\quad &{\rm {for \quad boson}}\cr
 {\pi (2n+1)}{i T}  \quad &{\rm {for \quad fermion}},\cr}
\cr}
$$
where $n$ runs over $ 0,\pm 1 , \pm 2 \cdots $.
Then one-loop contributions at finite temperature can be written as
\eqn\am{\eqalign{
I_B(a)&={{T^4}\over {2 \pi^2 }} \int_0^{\infty} dx \,\,  x^2\,
{\rm ln}(1-{\rm e}^{-(x^2+a^2)^{1/2}}), \cr
I_F(a)&={-{4T^4}\over {2 \pi^2 }} \int_0^{\infty} dx \,\, x^2\,
{\rm ln}(1+{\rm e}^{-(x^2+a^2)^{1/2}}). \cr
}
}
Here $a^2\equiv \ph^2M^2(v)/T^2 $ with $M(v)$ being a mass of the
relevant particle, and $B(F)$ stands for boson (fermion).
Thus the finite-temperature part of the effective potential is
\eqn\an{\eqalign{
\bar U^{T}(\ph,T) 
& =  6I_B({\ph M_W}/T)+3I_B({\ph M_Z}/T)  \cr
&+2I_B({\ph M_{H^{\pm}}} /T)+I_B({\ph M_A}/T)   \cr
&+I_B({\ph M_h}/T)+I_B(0) + 2I_B(0)        \cr
&+\sum_A\big[3 I_F({\ph M_A^{(u)}}/T)+3
I_F({\ph M_A^{(d)}}/T)+I_F({\ph M_A^{(l)}}/T)\big] \cr
&+16I_B(0)+{3 \over 2}I_F(0),  \cr
}
}
where we have set $N_g=N_c=3$. We have added $\ph$-independent
contributions for completeness : $16I_B(0)$ for gluons, $2I_B(0)$
for photon, $I_B(0)$ for scalon and $(3/2)I_F(0)$ for neutrinos. 
Gauge-dependent terms vanish in $\bar U^{T}$ as in $V_1$. For
numerical analysis, we ignore masses of fermions except for the
top quark mass, and normalize the potential $\bar U^T(\ph,T)$ as
\eqn\ao{
\bar V^T(\ph,T) \equiv \bar U^T(\ph,T)-\bar U^T(\ph=0,T).
}
Eventually the effective potential at finite temperature in
one-loop approximation is written as
\eqn\ap{
V(\ph,T)=V_1(\ph)+{\bar V}^T(\ph,T),
}
where $V_1$ is given by \ai\ and \aj\ , and ${\bar V}^T$ by \an\ and \ao.
\newsec{The Phase Transition}
\subsec{ The critical temperature}
In this section we investigate phase structure of our model,
and determine critical temperature $T_C$ by numerical analysis
for various sets of scalar masses including set I and set II.
In the case of the second-order phase transition, $T_C$ is
defined as a point at which the second derivative of the
finite-temperature effective potential at the origin
vanishes. On the other hand $T_C$ in the first-order transition
is a temperature at which there appear two degenerate minima
in the potential. It is obvious that the potential \ap\  is a
smooth real function of $\ph$  at any $T$.  We can, therefore,
find nature of phase transition by calculating \ap\  directly
without further approximations. Curves of the effective potential
for various temperatures are depicted in \fig\fa{The effective
potential for scalar mass set I. The critical temperature is
\hbox{67.5 GeV.}} ( mass parameter set I ), and \fig\fb{The
effective potential for scalar mass set II. The critical temperature
is \hbox{97.6 GeV.}} ( mass parameter set II ). These figures show
clearly that the phase transition is of first order. We find
$T_C\simeq 67.5$ GeV for set I and $T_C\simeq 97.6$ GeV for set II.
The critical temperatures for various sets of scalar masses 
are given in table 1.  We see that $\ph M_{\rm max}(v)/T_C > 1$ for
$\ph \sim \ph_+$, where $M_{\rm max}(v)$ is the heaviest scalar
mass and $\ph_+\,(\not = 0)$ the order parameter at a local minimum.
That is, high-temperature expansion cannot be adopted in a parameter
region of interest.\par
In the standard model, it is known that the condition not to
wash out baryon density after EWPT is roughly expressed as
\ref\brm{D.~E.~Brahm, Proceedings of the XXVI International
Conference on High Energy Physics (1993) 1583,  and
references therein.}
\eqn\aoo{
\phi_+(T_C)/T_C > 1.4 .
}
Here $\phi_+(T_C)$ is written as $v\ph_+(T_C)$ in our notation.
One can see from \fa\  and \fb\  that  $\phi_+(T_C)/T_C$  is large
due to low $T_C$ and large $\ph_+(T_C)\sim 1$. The inequality \aoo\  is
satisfied for almost all cases in our model. Thus, if the sphaleron
mass is the same order as in the single-doublet case, generated or
primordial baryon number is preserved after EWPT.  Of course the
sphaleron solution does not exist at the classical level in the
massless-doublet model.  It is conceivable in principle, however,
that the solution exists if one takes into account the one-loop
potential $V_1(\ph)$, because a mass scale appears as a
renormalization scale or as the VEV $v$.
\subsec{ The critical bubble}
It is well known that the bubble is nucleated in symmetric
phase when the phase transition is of first order. The rate of
bubble nucleation per unit volume is given by a formula\brm\ :
\eqn\aaa{
\Gamma_b=T^4 \Big({ F\over {2 \pi T}}\Big)^{3/2}{\rm e}^{-F/T},
}
where $F$ is the three-dimensional $O(3)$-invariant
bubble action given by
\eqn\aab
{
F=\int d^3{\mib x}\Big[{1\over 2}(\nabla \phi)^2+V(\phi,T)\Big].
}
In order to convert the Universe into the broken phase,
the bubble nucleation rate $\Gamma_b$ must at least be larger than
that of the expansion of the Universe, $\Gamma_H \sim T/m_{pl}$.
The dominant contribution to $\Gamma_b$ is given by minimizing $F$.
The critical bubble is the one which minimizes $F$. So we must compute
the critical bubble action. To compute the critical bubble, we must
solve the Euler-Lagrange equation given by
\eqn\aac{
-{{d^2 \phi}\over {d r^2}}-{2 \over r}{{d \phi}\over{ d r}}+{{\del
V}\over
{\del
\phi}}=0. }
Substituting the solution of \aac\  back into \aab\ , we obtain the
free energy $F_C$ of the critical bubble. But in general it is very
difficult to solve \aac\  analytically. In particular, in our case
$V$ is given by the integral form so that even a numerical analysis
is complicated. Instead of solving \aac\ , we make some approximations
to obtain the critical bubble. Given a potential as in
\fig\fc{The effective potential below the critical temperature.},
which shows a potential below the critical temperature,
we first minimize $F$ with respect to shape\brm. It is expected
that the solution takes a spherically symmetric shape. Then $F$ can be
approximately written as
\eqn\aad{
F \simeq 4 \pi R^2 \delta \Big({1\over 2}({\phi_{+}
\over \delta})^2 +V_{max}\Big)
-{{4 \pi}\over 3} R^3 \epsilon,
}
where $R$ is the radius of the bubble and $\delta$ the thickness
of the bubble wall.  The depth $\epsilon$ ($>0$) at the absolute
minimum, $V_{max}$ and $\phi_+$ are defined in \fc\ .
Next we extremize $F$ with respect to $\delta$  and $R$ to obtain
\eqn\aae{
\delta^2={ {\phi_{+}}^2 \over {2V_{max}}}, \quad
       R={{2 \phi_{+}} \over \epsilon} \sqrt{2V_{max}}.
}
Thus the free energy of the critical bubble $F_C$ is given by
\eqn\aaf{
F_C={{16\pi}\over {3\epsilon^2}}\Big(\phi_{+} \sqrt{2V_{max}}\Big)^3 .
}
In \fig\fd{The free energy of the critical bubble divided
by $T$ for set I vs. temperature in GeV unit.} and
\fig\fe{The free energy of the critical bubble divided
by $T$ for set II vs. temperature in GeV unit.} we
plot $F_C/T$ against temperature for  scalar-mass sets I and II,
respectively. The criterion,
$\Gamma_C > \Gamma_H$ implies roughly $F_C /T < 145$,
where $\Gamma_C\equiv \Gamma_b(F=F_C)$.
At temperatures below the dotted lines drawn in \fd\  and \fe, the
bubble can nucleate in the symmetric phase and the Universe can be
converted into the broken phase. The ``critical'' temperatures at
which the bubble can nucleate are about $52.5$ GeV and  $80.5$ GeV
for sets I and II, respectively. These temperatures are close to
the critical temperature for each case.
\newsec{Conclusions and Discussions}
We have studied the one-loop effective potential of electroweak theory
with two massless Higgs doubles at finite temperature. We obtained
the effective potential for various temperatures for two sets of
scalar masses without high-temperature expansion. We have found
that the phase transition in this model is of first order. We also
calculated the free energy of the critical bubble with some
approximations and found that the bubble nucleation can occur a
little below the critical temperature. \par
In order to generate baryon number at the EWPT, it is necessary for
the model to satisfy the Zakharov's three conditions, (i), (ii) and
(iii). In our model the condition (iii) is satisfied as shown in
section 3. The condition (i) is obviously satisfied in our model
because the gauge structure of this model is the same as that of
one-Higgs-doublet model. Since there is no new CP violating phase
in the Higgs sector in our model, the original Kobayashi-Maskawa
phase is the only source of CP violation, though the condition
(ii) is satisfied. It seems that our model is very attractive to
generate baryon number at the EWPT. But there are some questions to be
answered. We discuss these points below. Detailed analysis will be
given in our forthcoming paper. \par
The first question is how we understand the sphaleron induced anomalous
process in the broken phase in this model. The sphaleron is a solution
of classical equations of motion in electroweak theory. Its mass
(sphaleron energy) gives the height of energy barrier between
topologically inequivalent vacua. If there are no tree-level mass
terms of the Higgs fields, which corresponds to our model, no such
kinds of solution exist. That is, no sphaleron solution exists in our
model at the classical level. We expect, however, the solution may exist
if we consider $V_1(\ph)$, as mentioned in the previous section. The
sphaleron solution in two-Higgs-doublet model with tree-level mass
terms was studied in ref.\nref\kpz{B.~Kastning, R.~D.~Peccei
and X.~Zhang, \PL{266} (1991) 413.}\kpz\ by Kastning {\it et al}\/.
{}From their analysis we see that the dependence of the sphaleron
energy on the tree-level mass terms is not so significant and that
the energy mainly depends on the VEV of the Higgs fields and on the
gauge coupling. So it is quite possible that the rate of the anomalous
process in this model in the broken phase may be estimated by using
the result of the sphaleron solution in the model with tree-level mass
terms. The second question is whether observed baryon asymmetry can be
realized or not. In our model rather strong first-order phase transition
occurs and the latent heat is released after the phase transition.
This latent heat is converted into the entropy factor which has much
effect on baryon asymmetry. If the entropy is very large, there is a
possibility to wash out produced baryon asymmetry. Fortunately, we
found in section 3 that the ``critical'' temperatures at which the
bubble begins to nucleate are very close to the critical temperatures.
Therefore we expect that the entropy generation is not so significant
that baryon density may not be washed out, because the period of
supercooling is short. It is also an important problem to clarify
how the magnitude of the CP violation in the standard model affects
on the baryon number in our model. In addition to above two questions,
we must analyze many problems to confirm scenarios of electroweak
baryogenesis. Specifically, detailed understanding of the dynamics
of the EWPT must be developed. Understanding the nucleation of the
bubbles of the broken-symmetry phase is one of the most important
subjects. At this stage the bubble nucleation and its growth are not
so well-understood even in the standard model.\par
There are many open problems to be studied in two massless
Higgs-doublet model, but we believe that this model can provide
many attractive features for electroweak baryogenesis.
%
%
\bigbreak\bigskip
\centerline{{\bf Acknowledgements}}\nobreak
The authors would like to express their cordial gratitude to
S.~Otsuki, F.~Toyoda and other colleagues at
Saga, Kyushu and Kinki Universities for discussions and encouragement.
One of the authors (K.F.) is partially supported by the Grant-in-Aid
for Encouragement of Young Scientist of the Ministry of
Education, Science and Culture (No. 05740182).
\bigskip\bigskip\footatend\vfill\immediate\closeout%
\rfile\writestoppt\baselineskip=14pt\centerline{{\bf References}}%
\nobreak\bigskip{\frenchspacing%
\parindent=20pt\escapechar=` \input refs.tmp\vfill}\nonfrenchspacing
\vfill
\bigskip\bigskip\vfill{\parindent40pt
\baselineskip14pt\centerline{{\bf Table Caption}}\nobreak\medskip
\item{Table\ 1.}The critical temperatures for various sets of scalar masses.
We assume $M_t=140$ GeV. GeV unit is used.\vfill}
\bigskip\bigskip\vfill\immediate\closeout%
\ffile{\parindent40pt\baselineskip14pt\centerline%
{{\bf Figure Captions}}\nobreak\medskip
\escapechar=` \input figs.tmp\vfill\eject}
%
\centerline{}
\vskip 26ex
\begintable
 $M_{H^{\pm}}$ | $M_A$ | $M_h$ | $M_S$       | $T_C$     \cr
  70           | 350   | 350   |   73.57     | 65.7      \cr
  150          | 350   | 350   |   74.93     | 65.0      \cr
  150          | 400   | 250   |   74.22     | 64.8      \cr
  250          | 350   |  70   |   62.49     | 57.9      \cr
  250          | 350   | 150   |   63.29     | 57.8      \cr
  250          | 400   | 400   |  107.06     | 81.2      \cr
  350          | 70    | 350   |   92.41     | 74.8      \cr
  350          | 250   | 200   |   80.95     | 67.5      \cr
  350          | 400   | 400   |  126.88     | 89.9      \cr
  400          | 250   | 250   |  107.06     | 81.1      \cr
  400          | 350   | 350   |  126.88     | 89.9      \cr
  450          | 350   | 250   |  142.19     | 97.6      \endtable
\smallskip
\centerline{\bf Table 1}
\vfill\eject
\nopagenumbers
\epsfxsize=0.86\hsbody
\centerline{\epsfbox{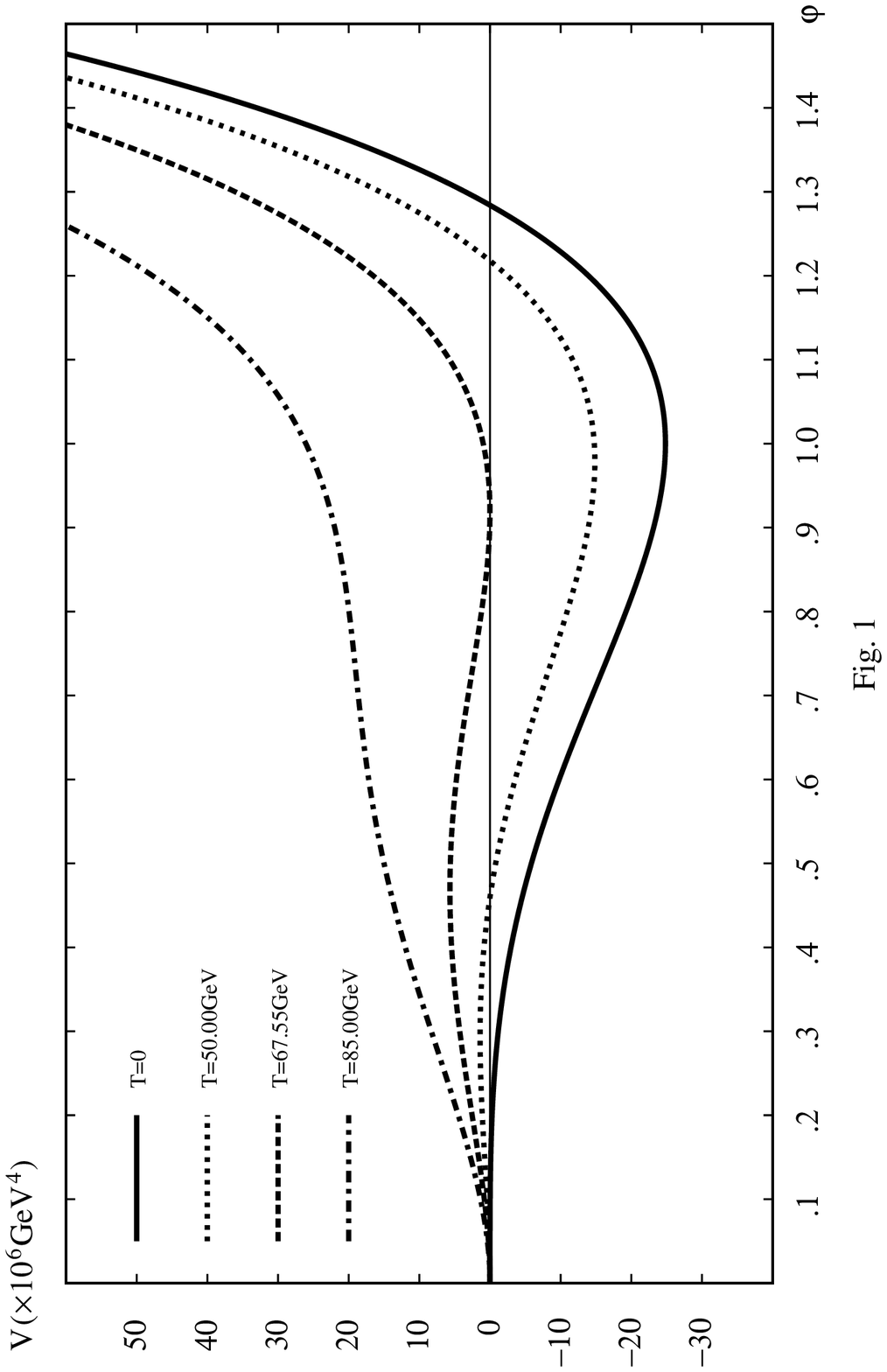}}
\vfil\eject
\nopagenumbers
\epsfxsize=0.86\hsbody
\centerline{\epsfbox{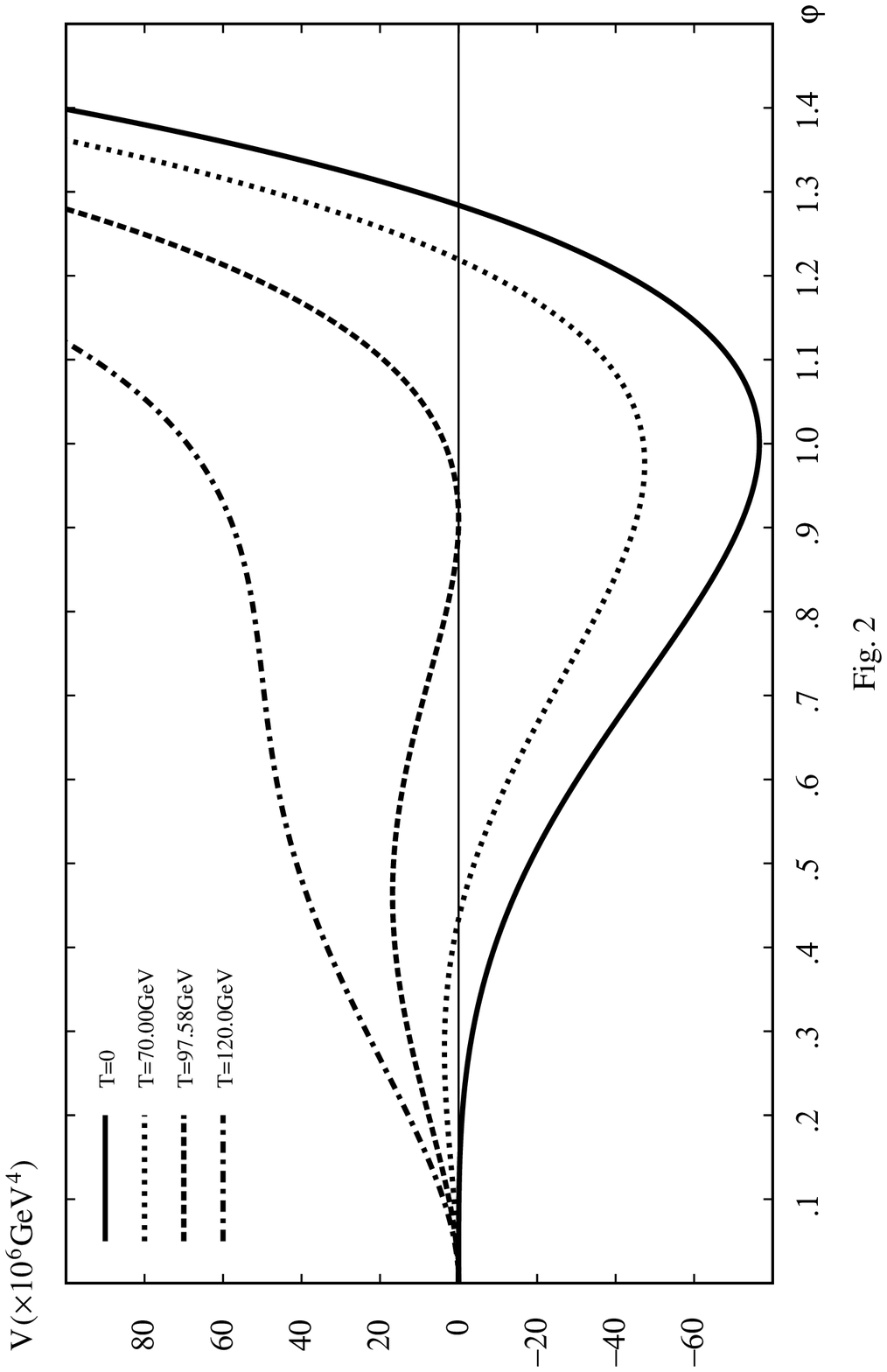}}
\vfil\eject
\nopagenumbers
\epsfxsize=0.86\hsbody
\centerline{\epsfbox{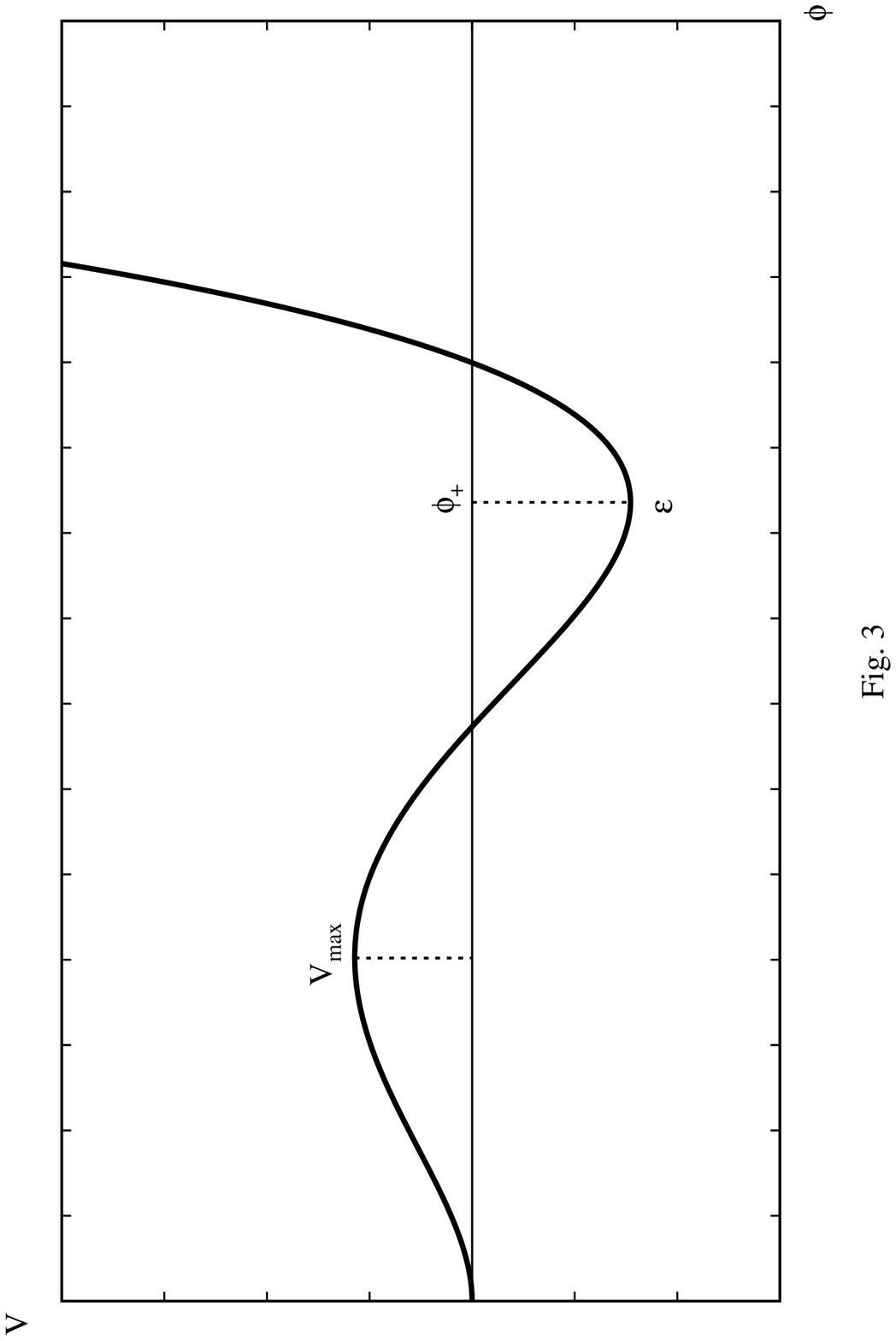}}
\vfil\eject
\nopagenumbers
\epsfxsize=0.86\hsbody
\centerline{\epsfbox{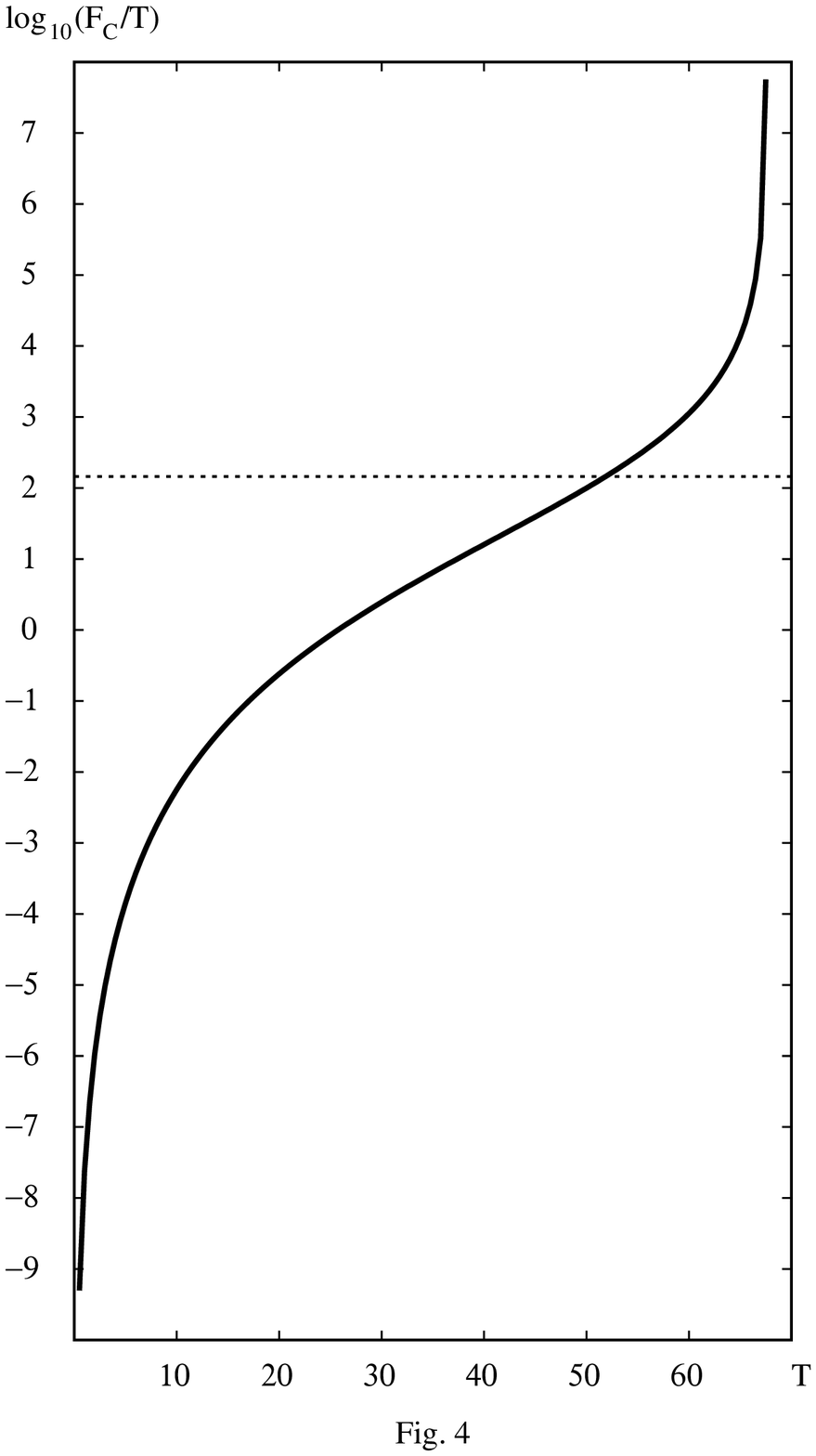}}
\vfil\eject
\nopagenumbers
\epsfxsize=0.86\hsbody
\centerline{\epsfbox{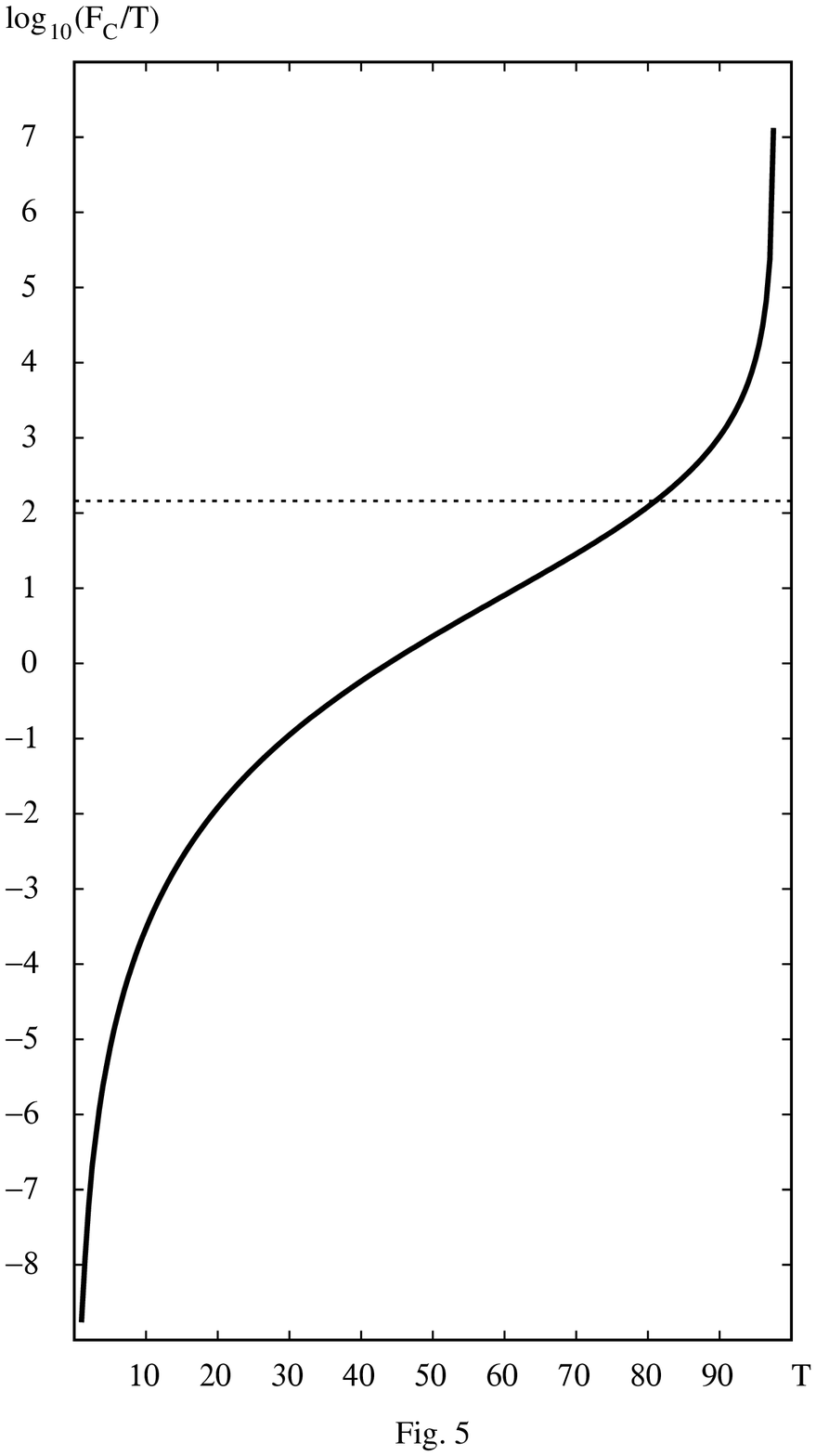}}
\vfil\eject
\bye